\documentclass[pra,twocolumn,aps,showpacs,amssymb,superscriptaddress]{revtex4}
\usepackage{graphicx}

\begin{document}

\title{Two characteristic temperatures for a Bose-Einstein condensate of a finite number of particles}

\author{Z. Idziaszek}
\affiliation{Centrum Fizyki Teoretycznej, Polska Akademia Nauk, 02-668 Warsaw, Poland}
\affiliation{Institut f\"ur Theoretische Physik, Universit\"at Hannover, D-30167 Hannover,Germany}
\author{K. Rz\c{a}\.{z}ewski}
\affiliation{Centrum Fizyki Teoretycznej, Polska Akademia Nauk, 02-668 Warsaw, Poland}
\affiliation{Dept. of Physics and Astronomy, University of Rochester, NY 14627, USA}

\begin{abstract}
We consider two characteristic temperatures for a Bose-Einstein condensate, 
that are related to certain properties of the condensate statistics.
We calculate them for an ideal gas confined in power-law traps
and show that they approach the critical temperature in the limit of large number of particles.
The considered characteristic temperatures can be useful in the studies of Bose-Einstein condensates
of a finite number of atoms, indicating the point of a phase transition.
\end{abstract}


\maketitle


Experimental achievement of Bose-Einstein condensation (BEC) in trapped, cold gases of 
alkali atoms \cite{BEC}, has stimulated large interest in the physics of this phenomenon.
Among others, the issue of fluctuations in the number of condensed atoms has been a subject of 
intensive theoretical studies, both for noninteracting 
\cite{Politzer,Wilkens,SPcalc,SPGross,MaxDem,MDGross,MDWeiss,Schnack,Holthaus} and interacting gases 
\cite{FluktInt}. 
For an ideal Bose gas, described in the canonical or microcanonical ensemble, 
the condensate fluctuations are maximal just below the critical temperature. 
The point of maximal fluctuations defines some characteristic temperature, which can indicate the occurence of 
a phase-transition \cite{Schnack}. In the systems containing finite number of particles, the thermodynamic functions 
remain analytic for all temperatures, and definition of the critical temperature cannot be based on the presence 
of a singularity in the thermodynamic functions. In this case, the characteristic temperature provides an 
alternative way to localize the point of a phase-transition. 
 
Different characteristic temperature can be defined on the basis of the ground-state occupation number. 
Mean ground-state population as a function of the temperature, exhibits the presence of the inflexion point close 
to the critical temperature. In this paper we study the properties of both characteristic temperatures, 
demonstrating that they tend to the critical temperature in the limit of large number of particles.

The definition of a suitable characteristic temperature for finite 
systems was already considered by several authors.
One of the candidates was a cross-over temperature. It is the temperature 
at which the probability distribution of the occupation of the 
condensed state changes its character from the one with the maximum at 
zero atoms (above phase transition) to the one with the maximum at 
nonzero (below the phase transition) \cite{MDGross,MDWeiss,Kocharovsky2,WilkensInt}. 
Another candidate was the maximum in the specific heat \cite{Kirsten,Haugerud,Pathria}.
We stress that there are other characteristic temperatures possible, all tending to the critical one 
in the limit of $N$ going to infinity and, perhaps, one of them may be 
less difficult to measure in the experiment or compute rigorously for 
the interacting system.


We describe the system of $N$ noninteracting bosons, at a temperature $T$, using the canonical ensemble.
The canonical partition function $Z(N,T)$, is, by definition related to 
grand canonical one $\Xi(z,T)$, through equation:
$\Xi(z,T)=\sum_{N=0}^{\infty} z^N Z(N,T)$. The grand canonical 
partition function is known in the compact form:
$\Xi(z,T)= \prod_{\nu=0}^{\infty} (1-z e^{-\beta \varepsilon_{\nu}})^{-1}$ where $\beta=1/(k_B T)$,
and $\varepsilon_{\nu}$ is the single-particle energy of the level $\nu$.
Unfortunately, no such general expression exists for $Z(N,T)$, and we calculate $Z(N,T)$ from $\Xi(z,T)$,
using Cauchy integral formula 
\begin{equation}
\label{ZnInt}
Z(N,T) = \oint \frac{dz}{2 \pi i} \frac{\Xi(z,T)}{z^{N+1}}.
\end{equation}
In the similar manner, it is possible to express the ground-state occupation number
$\langle N_0 \rangle$ in the canonical ensemble 
\begin{equation}
\label{N0Int}
\langle N_0 \rangle = \frac{1}{Z(N,T)} \oint \frac{dz}{2 \pi i} \frac{\Xi(z,T)}{z^{N+1}} 
\frac{z}{1-z},
\end{equation}
and its fluctuations $\langle \delta^2 N_0 \rangle$   
\begin{eqnarray}
\label{dN0Def}
\langle \delta^2 N_0 \rangle & = & \langle N_0^2 - N_0 \rangle + 
\langle N_0 \rangle - \langle N_0 \rangle^2, \\
\label{dN0Int}
\langle N_0^2- N_0 \rangle & = & \frac{2}{Z(N,T)} \oint \frac{dz}{2 \pi i} 
\frac{\Xi(z,T)}{z^{N+1}} \left( \frac{z}{1-z} \right)^2.
\end{eqnarray}
Standard way of calculating the contour integrals (\ref{ZnInt}), (\ref{N0Int}) and (\ref{dN0Int}) is to use  
the saddle point (SP) approximation. This method exploits the fact that for large number of atoms the integrand 
is sharply peaked around the most probable value, and can be approximated by the Gaussian function. 
This approximation, however, ceases to be valid below the critical temperature, $T_C$, of the BEC, when
the saddle point lies close to the ground-state singularity. In this case SP method predicts improper 
results for the condensate fluctuations at low temperatures \cite{SPcalc,SPGross,Holthaus}
Alternatively, the Maxwell Demon (MD) ensemble \cite{MaxDem} can be used, in which the  
moments of the condensate statistics can be easily elaborated from the grand canonical partition 
function of excited subsystem. This approach, however, is not suitable for studying the condensate statistics in the 
transition region, and in the case of fluctuations, it does not predict the occurence of the maximum 
\cite{MaxDem,MDGross,MDWeiss}.

In this paper we apply a variant of the SP method, developed quite recently by Holthaus and Kalinovski \cite{Holthaus}. 
Following the idea of Dingle \cite{Dingle}, the authors of \cite{Holthaus} exclude the ground-state term from 
the Taylor expansion around the SP and perform further calculation with the special treatment of the ground-state term. 
In this way, they calculate mean number of condensed atoms and its fluctuations, in the whole range 
of temperatures, including the most interesting regime of the phase transition.
According to \cite{Holthaus} we rewrite Eq. (\ref{ZnInt}) in the form
\begin{equation}
\label{ZnexpF}
Z(N,T) = \oint \frac{dz}{2 \pi i} \frac{\exp(-F(z,T))}{1-z},
\end{equation}
where $F(z,T) = (N+1) \ln z + \sum_{\nu=1}^{\infty} \ln (1 - z e^{- \beta \varepsilon_{\nu}})$ is a tempered 
function, with excluded contribution of the ground-state term. 
Function $F(z,T)$ should be expanded around the 
saddle point $z_0$, which is calculated for the whole integrand. It fulfills the following equation
\begin{equation}
\label{Sp0}
\frac{z_0}{1-z_0} + \sum_{\nu=1}^{\infty}
\frac{ z_0 e^{- \beta \varepsilon_{\nu}}}{1 - z_0 e^{- \beta \varepsilon_{\nu}}} = N + 1.
\end{equation}
Expansion of $F(z,T)$ around $z_0$ up to the second-order terms, leads to an integral involving a product of 
the Gaussian function with the singular ground-state term $1/(1-z)$. The result of integration may be expressed 
in terms of the parabolic cylinder function \cite{Holthaus}. In the regime of condensation, it can 
be approximated by relatively simple expression
\begin{equation}
\label{ZnSp}
Z(N,\beta) = e^{- F(z_0,T) - 1}.
\end{equation}
The same procedure may be repeated for the contour integrals (\ref{N0Int}) and (\ref{dN0Int}),
describing $\langle N_0 \rangle$ and $\langle \delta^2 N_0 \rangle$, respectively. 
The SP of the former integral, $z_1$, fulfills the following equation 
\begin{equation}
\label{Sp1}
2 \frac{z_1}{1-z_1} + \sum_{\nu=1}^{\infty}
 \frac{ z_1 e^{- \beta \varepsilon_{\nu}}}{1 - z_1 e^{- \beta \varepsilon_{\nu}}} = N,
\end{equation}
while the SP of the latter integral, $z_1$, has to be determined from 
\begin{equation}
\label{Sp2}
3 \frac{z_2}{1-z_2} + \sum_{\nu=1}^{\infty}
 \frac{ z_2 e^{- \beta \varepsilon_{\nu}}}{1 - z_2 e^{- \beta \varepsilon_{\nu}}} = N - 1.
\end{equation}
Similar procedure, as that used for derivation of (\ref{ZnSp}), applied to the integrals  (\ref{N0Int}) and (\ref{dN0Int}),
yields the following results \cite{Holthaus} 
\begin{eqnarray}
\label{n0Sp}
\langle N_0 \rangle & = & 2 \frac{z_1}{1-z_1} e^{ F(z_0,T) - F(z_1,T)-1}, \\
\label{dn0Sp}
\langle N_0^2 - N_0 \rangle & = & 9 \left( \frac{z_2}{1-z_2} \right)^2
e^{F(z_0,T) - F(z_2,T)-2}.
\end{eqnarray}
Eqs. (\ref{n0Sp})-(\ref{dn0Sp}), with the saddle points $z_0$, $z_1$ and $z_2$ calculated from
Eqs. (\ref{Sp0}), (\ref{Sp1}), and (\ref{Sp2}) respectively, determine the mean ground-state
occupation number 
$\langle N_0 \rangle$ and the fluctuations $\langle \delta^2 N_0 \rangle$ for temperatures $T<T_C$.
Comparison with the exact numerical results calculated for moderate-size systems 
containing $10^2$ - $10^4$ atoms demonstrate, that the modified SP method predicts very accurate values of
$\langle N_0 \rangle$ and $\langle \delta^2 N_0 \rangle$ .
The accuracy of the SP approximation grows with the number of atoms, which reflects the fact, 
that for larger $N$ the integrand becomes more sharply peaked in the SP.

Derivation of the analytical results for the considered characteristic temperatures, requires to 
solve, at least in some approximation, the implicit equations determining the saddle points. 
The approximate solutions of  Eqs. (\ref{Sp0}), (\ref{Sp1}) 
and (\ref{Sp2}) may be found, by expanding their left hand side around  $z=1$, without altering the ground-state
term $z/(1-z)$. Applying this procedure to Eq. (\ref{Sp0}), we obtain
\begin{equation}
\label{SPapp0}
\frac{z_0}{1-z_0} + n_e(T) + d_e(T) (z_0 - 1) \simeq N + 1, \\ 
\end{equation}
where
\begin{equation}
\label{neDef}
n_e(T) = 
\sum_{\nu=0}^{\infty} \frac{ e^{- \beta \varepsilon_{\nu}}}{1 - e^{- \beta \varepsilon_{\nu}}}, 
\end{equation}
and 
\begin{equation}
\label{deDef}
d_e(T) = 
\sum_{\nu=0}^{\infty} \frac{ e^{- \beta \varepsilon_{\nu}}}{1 - e^{- \beta \varepsilon_{\nu}}} 
\left( \frac{ e^{- \beta \varepsilon_{\nu}}}{1 - e^{- \beta \varepsilon_{\nu}}} + 1 \right).
\end{equation}
After expansion, the SP Eq. (\ref{SPapp0}) becomes quadratic in  $z_0$. Its solution reads
\begin{equation}
\label{z0}
z_0 = \frac{n_0 + 2 d_e - \sqrt{n_0^2 + 4 d_e}}{ 2 d_e}, \\
\end{equation}
where $n_0(T)= N - n_e(T) + 2$. The same procedure repeated for the two remaining SP
Eqs. (\ref{Sp1}) and (\ref{Sp2}), leads to the following solutions 
\begin{eqnarray}
\label{z1}
z_1 & = & \frac{n_0 + 2 d_e - \sqrt{n_0^2 + 8 d_e}}{ 2 d_e}, \\
\label{z2}
z_2 & = & \frac{n_0 + 2 d_e - \sqrt{n_0^2 + 12 d_e}}{ 2 d_e}.
\end{eqnarray}
The distance of the saddle points $z_1$ and $z_2$ to $z_0$, in the complex plane, is very small,
which may be easily seen from Eqs. (\ref{z0}), (\ref{z1}), (\ref{z2}). Hence, $F(z_1,T)$ and 
$F(z_2,T)$, appearing in Eqs. (\ref{n0Sp}) and (\ref{dn0Sp}), may be calculated by means of the 
Taylor expansion around $z=z_0$. This will allow to further
simplify Eqs. (\ref{n0Sp}) and (\ref{dn0Sp}). In order to 
preserve sufficient accuracy required for calculation of the characteristic temperature, we perform expansion up to the 
second order terms in the small parameters $(z_1-z_0)$ and $(z_2-z_0)$ for $F(z_1,T)$ and $F(z_2,T)$, respectively.
In the case of $F(z_1,T)$, we obtain  
\begin{eqnarray}
F(z_1,T) & \simeq & F(z_0,T) + \frac{z_1 -z_0}{1 -z_0} - \nonumber \\
\label{Fexp}
& - & \frac{1}{2}\left( d_e(T) + \frac{z_0}{1 -z_0}
\right) \left( \frac{z_2 - z_0}{z_0} \right)^2,
\end{eqnarray}
where we have applied the following approximation
$\frac{\partial^2 F}{\partial z^2} (z_0) \simeq -d_e(T) - z_0/(1-z_0)$. The small error 
that results from this approximation is, in fact, a higher order term, which may be safely neglected.
In the same way we expand $F(z_2,T)$, and substitute the expanded functions into (\ref{n0Sp}) and (\ref{dn0Sp}), 
obtaining
\begin{eqnarray}
\label{n0Spa}
\langle N_0 \rangle & = & 2 \frac{z_1}{1-z_1} \times \nonumber \\ 
& \times & e^{
\left( \frac{z_0 - z_1}{z_0} \right)^2 \left[ \frac{z_0}{1-z_0} \left( \frac{1}{2} + \frac{z_0}{z_0-z_1}
\right) + \frac{1}{2} d_e(T) \right] - 1} ,\\
\label{dn0Spa}
\langle N_0^2- N_0 \rangle & = & 9 \left( \frac{z_2}{1-z_2} \right)^2 \times \nonumber \\ 
& \times & e^{ \left( \frac{z_0 - z_2}{z_0} \right)^2 \left[ 
\frac{z_0}{1-z_0} \left( \frac{1}{2} + \frac{z_0}{z_0-z_2} \right) + \frac{1}{2} d_e(T) \right] - 
2 }.\makebox[1cm]{}
\end{eqnarray}  
\begin{figure}[tt]
\includegraphics[width=6cm,clip]{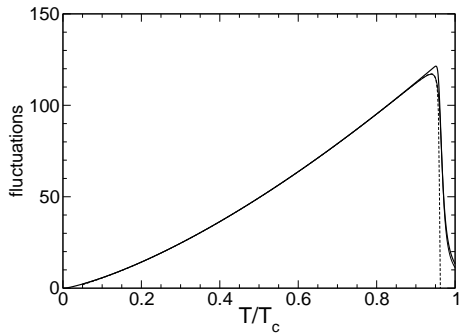}
\caption{
\label{fig:1}
Canonical fluctuations of the condensate occupation number for $N=10^4$ atoms. 
Thick solid line depicts the exact numerical result. Thin solid line corresponds to the numerical 
result calculated from Eqs. (\ref{n0Sp}) and (\ref{dn0Sp}) derived in the SP approximation. 
Dashed lines represents the analytical result (\ref{dN0Max}).}
\end{figure}
We have performed several numerical tests, which proved that Eqs. (\ref{n0Spa})-(\ref{dn0Spa}), with the 
saddle points given by (\ref{z0})-(\ref{z2}), predict very accurate values of 
$\langle N_0 \rangle$ and $\langle \delta^2 N_0 \rangle$. In comparison to the initial Eqs. (\ref{n0Sp}) and (\ref{dn0Sp}),
they do not require numerical solving of equations for the saddle points. 
Nevertheless, developed set of equations is still too complicated, for the analytical calculation of the 
characteristic temperature. To this end, we expand Eqs. (\ref{z0})-(\ref{z2}) and 
(\ref{n0Spa})-(\ref{dn0Spa}) in the power series in small parameter $N^{-1}$, assuming that 
$n_e \sim N$ and $d_e \sim N$. After straightforward, but tedious calculations we arrive at 
\begin{eqnarray}
\label{N0Infl}
\langle N_0 \rangle & = & n_0 + \frac{1}{2} \frac{d_e}{n_0} -\frac{3}{2}, \\
\label{dN0Max}
\langle \delta^2 N_0 \rangle & = & d_e - \frac{3}{2} \left( \frac{d_e}{n_0} \right)^2 
- \frac{3}{2} \frac{d_e}{n_0} - 2,
\end{eqnarray}
where in the final result we include the terms of the order of $O(1)$, and higher. 
In the thermodynamic limit, only the leading order terms are important: $n_0$ in Eq. (\ref{N0Infl}) and $d_e$ in 
Eq. (\ref{dN0Max}). Those terms can be derived from the MD ensemble \cite{MaxDem}. 
The inclusion of the remaining terms is, however, crucial for the occurrence of the inflexion point in 
$\langle N_0 \rangle$, and the maximum in $\langle \delta^2 N_0 \rangle$.

Figure~\ref{fig:1} shows the fluctuations  $\langle \delta^2 N_0 \rangle$ for an ideal gas of $N=10^4$ atoms, 
confined in 3D harmonic trap. The exact numerical result calculated by means of recurrence relations,
is compared with predictions of Eqs. (\ref{dn0Sp}) and (\ref{dN0Max}). In the case of Eq. (\ref{dn0Sp}), the 
saddle points $z_0$, $z_1$, and $z_2$, are calculated numerically from 
Eqs. (\ref{Sp0}), (\ref{Sp1}), and (\ref{Sp2}),
respectively. It is clearly seen that Eq. (\ref{dn0Sp}) obtained in the SP approximation, fits very well 
the exact
numerical curve, apart from the region close to the maximum, where a small discrepancy is present. 
This difference, however, does not influence significantly the temperature of the maximal fluctuations.
On the other hand, in a wide range of temperatures Eqs.(\ref{dn0Sp}) and (\ref{dN0Max}) gives almost identical values of 
$\langle \delta^2 N_0 \rangle$. Their predictions are different in the vicinity of the critical temperature, 
where $n_0 \simeq 0$ and Eq.(\ref{dN0Max}) becomes invalid. It should be stressed, however, that this 
wrong behavior 
at $T_C$ does not influence the position of maximum, which remains exactly the same as in (\ref{dn0Sp}). 
\begin{figure}[tt]
\includegraphics[width=6cm,clip]{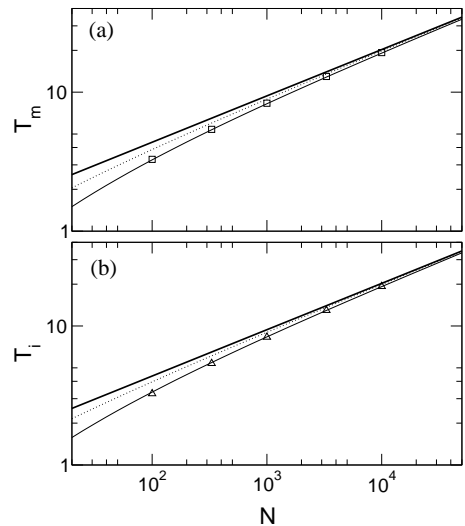}
\caption{
\label{fig:2}
Temperature of the maximal fluctuations $T_m$ (a) and the temperature of the inflexion point $T_i$
(b) as a function of the number of atoms $N$, calculated in the canonical ensemble for the case of
3D harmonic trap. Thick solid line on both plots depicts the critical temperature calculated 
in thermodynamic limit. Figure (a): $T_m$ given by Eq. (\ref{TmHarm}) (dotted line), is compared with 
the analytical result accounting also for lower order terms (thin solid line), and with exact data 
of numerical calculations (squares). Figure (b): $T_i$ given by Eq. (\ref{TiHarm}) (dotted line) is compared 
with the analytical result accounting also for lower order terms (thin solid line), and with exact data of
numerical calculations (triangles).}
\end{figure}

Now, we turn to the calculation of the temperature of maximal fluctuations $T_m$ and the temperature 
of inflexion point $T_i$. We consider an ideal Bose gas confined in 
{\it D}-dimensional trap with power-law single-particle energy spectra. The condensate fluctuations
in such systems have been studied \cite{MDGross,MDWeiss}; in this paper we adopt the notation of 
\cite{MDWeiss}. The energy spectra is given by 
$\varepsilon_{\{\nu_i\}}=\Delta \sum_{i=1}^{D} \nu_{i}^{\sigma}$, where $\Delta$ 
is the energy gap between the ground-state and the first excited state, $\sigma$ is an 
exponent depending on the shape of potential, and $\nu_i$ are the integer numbers counting the excitations
in $i$-th cartesian direction. Here, we assume the case of spherically symmetric 
potential, stressing that our analysis can be extended for general case
of an anisotropic trap. Moreover, we are interested in the case $D>\sigma$, in which the condensation 
occurs below the critical temperature $T_C$ given by  
$k_{B} T_{C}/\Delta=(N/\zeta(d))^{1/d}\Gamma(1+1/\sigma)^{-\sigma}$ \cite{MDGross,MDWeiss}, 
where $d=D/\sigma$, and $\zeta(d)$ denotes Riemann's Zeta function. 
In power-law traps, functions $n_e(T)$ and $d_e(T)$ are given by~\cite{MDGross,MDWeiss}
\begin{eqnarray}
\label{nePowLaw}
n_e(T) = N \left( \frac{T}{T_c} \right)^d, \\
\label{dePowLaw}
d_e(T) = \alpha N \left( \frac{T}{T_c} \right)^{\gamma},
\end{eqnarray}
where we retain only the leading order terms.
In Eq. (\ref{dePowLaw}), the dimensionless coefficients $\gamma$ and $\alpha$ are given by 
$\alpha=\zeta(d-1)/\zeta(d)$, $\gamma=d$ in the case of $d>2$, while for $1<d<2$ they are given by 
$\alpha=\Gamma(1+1/\sigma)^{D-2\sigma} \zeta(3-d) \Gamma(d)^{-1} \zeta(d)^{-2/d}$, $\gamma=2$.
In the border case: $d=2$, $d_e(T)$ exhibits logarithmic dependence on $T$, however, we will not 
consider this special case. We substitute Eqs. (\ref{nePowLaw}) and (\ref{dePowLaw}) into 
Eqs. (\ref{N0Infl}) and (\ref{dN0Max}), and determine the characteristic temperatures 
from: $\frac{d \langle \delta^2 N_0 \rangle}{d T}|_{T=T_m}=0$ 
and $\frac{d^2 \langle N_0 \rangle}{d T^2}|_{T=T_i}=0$, obtaining 
\begin{eqnarray}
\label{TmPowLaw}
\frac{T_{m}}{T_c} & = & 1 - \frac{(3 \alpha)^{1/3}}{d} N^{\frac{(\gamma/d)-2}{3}},\\
\label{TiPowLaw}
\frac{T_{i}}{T_c} & = & 1 - \frac{1}{d} \left( \frac{\alpha d}{d-1}\right)^{1/3} N^{\frac{(\gamma/d)-2}{3}}.
\end{eqnarray}
It is worth stressing that the power-law depence on $N$ is exactly the same for $T_m$ and $T_i$. 
In the considered regime of parameters $\gamma/d <2$, and both characteristic temperatures 
approach $T_C$ when $N \rightarrow \infty$. In the specific case, of 3D harmonic potential 
($d=3$, $\gamma=3$), Eqs. (\ref{TmPowLaw}) and (\ref{TiPowLaw}) take form 
\begin{eqnarray}
\label{TmHarm}
\frac{T_{m}}{T_c} & = & 1 - \left(\frac{\zeta(2)}{9 \zeta(3)}\right)^{1/3} N^{-1/3}, \\
\label{TiHarm}
\frac{T_{i}}{T_c} & = & 1 - \left( \frac{\zeta(2)}{18 \zeta(3)} \right)^{1/3} N^{-1/3}. 
\end{eqnarray}

Figure~\ref{fig:2} shows the dependence of $T_m$ (upper plot) and $T_i$ (lower plot) 
on the number of atoms $N$, in the case of 3D harmonic trap. It compares the predictions of 
Eqs. (\ref{TmPowLaw}) and (\ref{TiPowLaw}) with exact results of numerical calculations. In addition 
it also presents the analytical result, derived with the inclusion of the lower order terms 
in $n_e(T)$ and $d_e(T)$. In this case we obtain the same $N^{-1/3}$ dependence of $T_m$ and $T_i$
as in Eqs. (\ref{TmHarm}) and (\ref{TiHarm}), but with a different prefactors equal to 
$\zeta(2)/2 \zeta(3)^{2/3}+(\zeta(2)/9 \zeta(3))^{1/3}$ and 
$\zeta(2)/2 \zeta(3)^{2/3}+(\zeta(2)/18 \zeta(3))^{1/3}$, respectively.
From Fig.~\ref{fig:2} we see that Eqs. (\ref{TmHarm}) and (\ref{TiHarm}) predicts slightly different 
values in comparison to the exact data. Nevertheless, the analytical curve accountinng also for lower 
order terms fits very well the numerical results.

In conclusion, we have studied two characteristic temperatures for a Bose-Einstein condensate,
defined by the point of maximal fluctuations in the ground-state occupation number
and by the inflexion point of the ground-state occupation number.
For a wide class of power-law traps, we have calculated both characteristic temperatures, showing 
that they approach the critical temperature in the limit of large number of particles.
The numerical calculations performed for 3D harmonic trap, reveal a good agreement with our 
analytical results. The concept of the characteristic temperature may be useful in the studies of 
the BEC with a finite number of atoms, where the characteristic temperature can indicate the 
occurence of a phase transition.


Z.I. acknowledges support from Polish KBN Grant No. 5-P03B-103-20 and the Alexander von Humboldt Stiftung.


\begin{thebibliography}{99}

\bibitem{BEC} M.H. Anderson {\it et al.}, Science {\bf 269}, 198 (1995); 
K.B. Davis {\it et al.}, Phys. Rev. Lett. {\bf 75}, 3969 (1995);
C.C. Bradley {\it et al.}, {\it ibid} {\bf 75}, 1687 (1995); {\bf 79}, 1170 (1997). 

\bibitem{Politzer} H.D. Politzer, Phys. Rev. A {\bf 54}, 5048 (1996).

\bibitem{Wilkens} M. Wilkens and C. Weiss, J. Mod. Opt. {\bf 44}, 1801 (1997).

\bibitem{SPcalc} M. Gajda and K. Rz\c{a}\.{z}ewski, Phys. Rev. Lett. {\bf 78}, 2686 (1997).

\bibitem{SPGross} S. Grossmann and M. Holthaus, Phys. Rev. Lett. {\bf 79}, 3557 (1997).

\bibitem{MaxDem} P. Navez {\it et al.},
Phys. Rev. Lett. {\bf 79}, 1789 (1997); 

\bibitem{MDGross} S. Grossmann and M. Holthaus, Opt. Ex. {\bf 1}, 262 (1997).
 
\bibitem{MDWeiss} C. Weiss and M. Wilkens, Opt. Ex. {\bf 1}, 272 (1997).

\bibitem{Schnack} H.-J. Schmidt and J. Schnack, Physica A {\bf 260}, 479 (1998).

\bibitem{Holthaus} M. Holthaus and E. Kalinowski, Ann. Phys. (N. Y.) {\bf 10}, 385 (1999). 

\bibitem{FluktInt} S. Giorgini, L. P. Pitaevskii, and S. Stringari, Phys. Rev. Lett. {\bf 80}, 5040 (1998);
Z. Idziaszek {\it et al.}, Phys. Rev. Lett. {\bf 82}, 4376 (1999);
V.V. Kocharovsky, Vl.V. Kocharovsky, and M.O. Scully, Phys. Rev. Lett. {\bf 84}, 2306 (2000);
H. Xiong {\it et al.}, Phys. Rev. A {\bf 65}, 033609 (2002). 

\bibitem{Kocharovsky2} V.V. Kocharovsky {\it et al.}, Phys. Rev. A {\bf 61} 023609 (2000).

\bibitem{WilkensInt} M. Wilkens {\it et al.}, J. Phys. B {\bf 33}, 779 (2000).

\bibitem{Kirsten} K. Kirsten and D.J. Toms, Phys. Rev. A {\bf 54}, 4188 (1996).

\bibitem{Haugerud} H. Haugerud, T. Haugest, and F. Ravndal, Phys. Lett. A {\bf 225}, 18 (1997).

\bibitem{Pathria} R.K. Pathria, Phys. Rev. A {\bf 58}, 1490 (1998).

\bibitem{Dingle} R.B. Dingle, {\em Asymptotic Expansions: Their Derivations and Interpretation} (Academic
Press, London and New York, 1973).

\end{thebibliography}
\end{document}